\documentclass[12pt,letterpaper]{article}

\usepackage[utf8]{inputenc}

\usepackage{latexsym}
\usepackage{verbatim}
\usepackage[a4paper, total={6.5in, 9in}]{geometry}
\usepackage{silence}
\usepackage{amsthm,amsmath,amsfonts,amssymb,verbatim, color}
\usepackage{graphicx}
\usepackage[percent]{overpic}
\usepackage{bm}
\usepackage{braket}
\usepackage{float}
\usepackage{mathtools}
\usepackage{appendix}
\usepackage[colorlinks=true,citecolor=blue,linkcolor=black,urlcolor=blue]{hyperref}
\usepackage{graphics} % Testing...
\graphicspath{ {figures/} }
\usepackage[caption=false]{subfig}

\usepackage[mathscr]{euscript}
\usepackage[export]{adjustbox}
\usepackage[bottom]{footmisc}
\usepackage{textcomp}
\usepackage{braket}

\title{\bf Quasi-integrability and nonlinear resonances in cold atoms under modulation }

\author{Rahul Gupta$^{1}$, Manan Jain$^{2}$, Sudhir R. Jain$^{3}$ \\ $^{1}$\textit{\small Department of Physics, Indian Institute of Technology, Mumbai 400076, India}\\
$^{2}$\textit{\small School of Physics and Astronomy, University of Birmingham} \\ \textit{\small Birmingham - B15 2TT, United Kingdom}\\
$^{3}$\textit{\small UM-DAE Centre for Excellence in Basic Sciences, University of Mumbai}\\ \textit{\small Vidyanagari Campus, Mumbai 400098, India}}

\begin{document}

\maketitle

\begin{abstract}
Quantum dynamics of a collection of atoms subjected to phase modulation has been carefully revisited. We present an exact analysis of the evolution of a two-level system (represented by a spinor) under the action of a time-dependent matrix Hamiltonian. The dynamics is shown to evolve on two coupled potential energy surfaces, one of them binding while the other one scattering type. The dynamics is shown to be quasi-integrable with nonlinear resonances. The bounded dynamics with intermittent scattering at random moments presents the scenario reminiscent to Anderson and dynamical localization. We believe that a careful analytical investigation of a multi-component system which is classically non-integrable is relevant to many other fields, including quantum computation with multi-qubit system.

\end{abstract}

\section{Introduction}

Evolution in the fields of ultracold atoms and quantum physics in the past few decades has led to recognition of these fields as a huge well-acclaimed arena for the exploration of popular subjects like quantum chaos \cite{PhysRevLett.73.2974}, Feshbach resonances\cite{thalhammer2008double,GiovPhD,cui2018broad,tanzi2018feshbach,d2007feshbach,marte2002feshbach,yurovsky2003three,vogels1997prediction,blackley2013feshbach,franzen2022observation,cho2013feshbach} and ultracold atomic mixtures\cite{catani2008degenerate,wacker2016universal,ferrari2002collisional,sawant2021thermalization,mosk2001mixture}, atom interferometry\cite{kasevich1991atomic,rosi2018detecting,rosi2017proposed,rosi2014precision,rosi2017quantum,jain2022reply,jain2021new,d2019measuring,salvi2018testing,tino2021testing,lamporesi2008determination,liu2019vortex,canuel2006six,aguilera2014ste,biedermann2015testing}, atomic clocks \cite{kale2022field,clockchapterbook,ludlow2015optical,barontini2022measuring,takamoto2005optical,borde2002atomic,ushijima2015cryogenic,derevianko2011colloquium,yamanaka2015frequency,zheng2022differential,poli2013optical}, quantum diffraction \cite{zhang1993quantum,jain2018classical} \& quantum thermodynamics\cite{Giov1,munoz2020dissipative,glatthard2022optimal,vogler2013thermodynamics}. This is due to rich internal structures, longer de Broglie wavelengths and tunable long-range interactions possessed by ultracold atoms. Furthermore, the research in the regime of lower temperatures has also been extended to molecules \cite{tim,vijayG}. Apart from these recent developments, there has been a sustained effort to realize parallels between atomic and condensed matter physics \cite{ozawa2019topological}. One of the ideas pursued with great interest is the localization of states in disordered systems, pioneered by Anderson \cite{anderson1958absence}. Due to a common sense analogy between disorder and chaos, a connection between localization of wavefunctions of classically chaotic systems, and, disordered lattices of infinite \cite{fgp} and finite extent \cite{jain1993} was brought out. Even in matter waves, the phenomenon of localization has been experimentally demonstrated \cite{billy2008direct}.  

Many years ago, an experiment carried out by the group led by Raizen \cite{PhysRevLett.73.2974} demonstrated the dynamical analog of Anderson Localization in a system of cold atoms. In this experiment, about one hundred thousand $^{23}$Na atoms were trapped in a spherical volume of 300 $\mu$m at a temperature of 17 $\mu$K. At the end of the preparation step, the temperature was turned off and a modulated standing light field was switched on for 10 $\mu$s. The Hamiltonian describing the interaction of the sodium atom with the light field is given by \cite{graham} 
\begin{equation}
H_0 = H_{\rm el} + \frac{p^2}{2m} + eF\cos\{ k_L[x - \Delta L \sin \omega t] \} \cos \omega_L t.
\end{equation}
Here, $H_{\rm el}$ contains the interaction of valence electrons with an atom. The last term denotes the electric dipole interaction of the electromagnetic field with an electron. Laser frequency and wavenumber are respectively denoted by $\omega _L$ and $k_L$, and $\omega$ is the modulation frequency. Standing waves are generated by directing two counter-propagating laser beams into the trap, and, the modulation is achieved by passing one beam through an electro-optical phase modulator. The beam is made to strike a mirror in a cavity of length $\Delta L$ which is moving with the modulation frequency, $\omega$. The laser frequency was chosen close to the D$_2$ line of sodium. The electronic Hamiltonian can be reduced to a two-level system containing ground state $\psi^{-}|g\rangle$ and an excited state, $\psi^{+}|e\rangle$.
\begin{equation}
\psi=\left( \begin{array}{c}
\psi^{+}\\
\psi^{-}\\
\end{array}\right)=\psi^{+}|e\rangle + \psi^{-}|g\rangle
\end{equation} 
Taking the energy average of the two states as zero energy, the matrix elements of $H_{\rm el}$ and $eF$ together give 
\begin{equation}
 H_{\rm el} + eF =
\begin{pmatrix}
 	\hbar\omega_0/2 & \hbar\Omega \\
	\hbar\Omega & -\hbar\omega_0/2 \\
\end{pmatrix} = \frac{\hbar\omega_0}{2}\sigma _z + \hbar\Omega \sigma _x 
\end{equation}
where the transition frequency is denoted by $\omega _0$, $\Omega$ denotes Rabi frequency, and $\sigma 's$ are the Pauli matrices. Thus, $H_0$ may be written as
\begin{alignat}{1}
H_0 = \frac{p^2}{2m}{\bf I} + \frac{\hbar\omega_0}{2}\sigma _z + &\hbar\Omega  \cos \{ k_L[x - \Delta L \sin \omega t] \} \cos (\omega _Lt) \sigma _x.    
\end{alignat}
where ${\bf I}$ denotes an identity matrix. 

After we present the general Hamiltonian below, in \S 2, we present the Hamiltonian under Rotating Wave Approximation. Within this approximation, the case of adiabatic perturbation for the two cases of small and large detuning is considered. In \S 3, the exact solution for this matrix Hamiltonian is given. The method transforms the dynamics under the matrix Hamiltonian to dynamics on potential energy surfaces. Classical dynamics reveals the presence of nonlinear resonances in \S 4. The classical system obeys the Kolmogorov-Arnold-Moser (KAM) theorem \cite{Lichtenberg1992}, and hence is quasi-integrable \cite{berry_aip}. In a related context of quantum Rabi model, a discussion on integrability \cite{braak} and symmetries \cite{braak_symm} has been presented relatively recently. 

Special solutions are discussed as they have been used to analyze experiments carried out by different groups. For each case discussed at the quantum mechanical level, we also present classical phase space pictures and show that this atomic system presents a very interesting and deep instance of the association of quasi-integrability and dynamical localization. The phase space pictures exhibit certain misleading features in the approximated Hamiltonian, compared to the exact Hamiltonian obtained by systematic expansion in powers of $\hbar$.\\

\noindent
{\it General Hamiltonian}\\

We now transform to a frame which is rotating with $\omega _L$ about the $z$-axis in spin space:
\begin{align}
\psi _{\rm rot} = \exp \left( i\omega _L\sigma _zt/2 \right) \psi.
\end{align}
Substituting $\psi$ in the Schr\"{o}dinger equation, $i\hbar \partial \psi/\partial t = H_0\psi$, we have the equation for the rotated wavefunction:
\begin{alignat}{1}
H_{\rm rot} = \frac{p^2}{2m}{\bf I} &+ \frac{\hbar (\omega_0 - \omega _L)}{2}\sigma _z + \hbar\Omega  \cos \{ k_L[x - \Delta L \sin \omega t] \}.\nonumber \\ &.\cos (\omega _Lt) e^{i\omega _L\sigma _z t/2}\sigma _x e^{-i\omega _L\sigma _z t/2}.
\end{alignat}
Using the standard identity, $e^{i\omega _L\sigma _z t/2}\sigma _x e^{-i\omega _L\sigma _z t/2} = \sigma _x \cos \omega _L t - \sigma _y \sin \omega _L t$, we have the transformed Hamiltonian:
\begin{alignat}{1}\label{eq:6}
H_{\rm rot} = \frac{p^2}{2m}{\bf I} &+ \frac{\hbar (\omega_0 - \omega _L)}{2}\sigma _z + \frac{\hbar\Omega}{2}  \cos \{ k_L[x - \Delta L \sin \omega t] \}.\nonumber \\ &.[\sigma _x (1 + \cos 2\omega_L t) - \sigma _y \sin 2\omega _Lt]. \end{alignat}
This is the general Hamiltonian for the physical situation described above where there are terms oscillating with twice the $\omega_L$. 

\section{Rotating Wave Approximation}

The Schr\"{o}dinger equation for $H_{\rm rot}$ is usually solved under the Rotating Wave Approximation (RWA) \cite{graham,stockmann_1999}. Here the terms oscillating with frequency $2\omega _L$ are neglected. This leads to a simplified Hamiltonian,
\begin{alignat}{1}
H_{rot}^{\rm RWA} &= \frac{p^2}{2m}{\bf I} + \hbar \Omega _{\rm eff} (\sigma _z\cos \alpha + \sigma _x \sin \alpha)
\end{alignat}
where 
\begin{alignat}{1}
\Omega _{\rm eff} &= \frac{1}{2} [(\omega _0 - \omega _L)^2 + \Omega ^2\cos ^2\{ k_L(x - \Delta L \sin \omega t)] \}]^{1/2}, \nonumber \\
\tan \alpha &= \frac{\Omega \cos[k_L(x - \Delta L \sin \omega t)]}{\omega _0 - \omega _L}.
\end{alignat}
Let us rotate the state of this Hamiltonian further in the spin space by an angle $(-\alpha/2)$ about the $y$-axis, to obtain a new state, $\psi'=\psi'^{+}|e\rangle + \psi'^{-}|g\rangle= \exp (i\alpha \sigma _y/2) \psi _{rot}$
\begin{equation}
\psi'=\left( \begin{array}{c}
\cos(\alpha/2)e^{i\omega_L t/2}\psi^{+} + \sin(\alpha/2)e^{-i\omega_L t/2}\psi^{-}\\
-\sin(\alpha/2)e^{i\omega_L t/2}\psi^{+} + \cos(\alpha/2)e^{-i\omega_L t/2}\psi^{-}\\
\end{array}\right)
\end{equation}
in which the second term is diagonal. Consequently, the equation satisfied by $\psi '$ is 
\begin{alignat}{1}
i\hbar \frac{\partial \psi '}{\partial t} = -\frac{\hbar}{2} \frac{\partial\alpha}{\partial t} \sigma_y\psi ' + e^{i\alpha\sigma_y /2} H_{\rm rot}^{\rm RWA}e^{-i\alpha \sigma_y /2}\psi ' 
= H_{\rm eff}^{\rm RWA}\psi '.
\end{alignat}
But this will transform the kinetic term as \cite{gba}:
\begin{alignat}{1}
&e^{i\alpha\sigma_y /2} p^{2}{\bf I}e^{-i\alpha \sigma_y /2}\psi ' = \left( p{\bf I}-\hbar{\bf A}\right)^{2}\psi '={\bf \Pi}^2 \psi ' \\
&{\bf A}=\frac{\sigma_y}{2}\frac{\partial\alpha}{\partial x}=\frac{-k_L \delta_{L}\Omega\sin[k_L(x-\Delta L \sin\omega t)]\sigma_y}{2\left( {\delta_L}^2 + \Omega^{2}\cos^{2}[k_L(x-\Delta L \sin\omega t)]\right)}
\label{A}
\end{alignat}
where ${\bf I}$ is an identity matrix. Now we can employ the well-known identity:
\begin{alignat}{1}
e^{i\alpha (\hat{n}.\vec{\sigma})}\vec{\sigma}e^{-i\alpha (\hat{n}.\vec{\sigma})} = \vec{\sigma} \cos 2\alpha + \hat{n} \times \vec{\sigma} \sin 2\alpha + \hat{n}(\hat{n}.\vec{\sigma}) (1 - \cos 2\alpha).
\end{alignat}
While the ``potential" part of the Hamiltonian becomes diagonal with these unitary transformations, the kinetic term modifies to $(p{\bf I} - \hbar {\bf A})^2$. This has terms of order 1, $\hbar$, and $\hbar ^2$ - thus, a semiclassical expansion (and not a perturbative expansion) appears naturally. Moreover, since ${\bf A}$ has non-zero diagonal matrix elements, there is a possibility of a geometric phase appearing in the state of the atoms as the system evolves. This is indeed due to the cavity modulation.
Dimensionally, $\hbar{\bf A}/e$ is a magnetic vector potential. $H_{\rm eff}^{\rm RWA}$ can be written as:
\begin{alignat}{4}
H_{\rm eff}^{\rm RWA}&=\frac{{\bf \Pi}^{2}}{2m} + \hbar\Omega_{\rm eff}\sigma_z - \frac{\hbar}{2}\frac{\partial\alpha}{\partial t}\sigma_y, \\
&=\left[ \frac{p^{2}}{2m} -\frac{\hbar^{2}}{4}\left( \frac{\partial\alpha}{\partial x} \right)^2\right]{\bf I} + \hbar\Omega_{\rm eff}\sigma_z  
+ \left( -\frac{\hbar}{2}\frac{\partial\alpha}{\partial t} -\hbar\frac{\partial\alpha}{\partial x}p + \frac{i\hbar^2}{2}\frac{\partial\alpha}{\partial x}\right)\sigma_y.
\end{alignat}
Except for terms of order O($\hbar ^2$), each of the terms can make a significant contribution.  
At this point, one of the possible simplifications occurs if $\alpha$ is slowly varying with time. This leads us to consider applying the adiabatic approximation, which we discuss now.

\subsection{Adiabatic variation}

We may neglect the term $\hbar\sigma_y d\alpha /dt$. But note that in this case:
\begin{equation}
\hbar\sigma_y \frac{d\alpha}{dt}=\hbar\frac{\partial\alpha}{\partial x}p\sigma_y + \hbar\frac{\partial\alpha}{\partial t}\sigma_y\rightarrow 0.
\end{equation}
The adiabatic Hamiltonian is:
\begin{alignat}{1}
H_{\rm ad}^{\rm RWA}=\left[ \frac{p^{2}}{2m} -\frac{\hbar^{2}}{4}\left( \frac{\partial\alpha}{\partial x} \right)^2\right]{\bf I} + \hbar\Omega_{\rm eff}\sigma_z 
+ \left( \frac{\hbar}{2}\frac{\partial\alpha}{\partial t} + \frac{i\hbar^2}{2}\frac{\partial\alpha}{\partial x}\right) \sigma_y.
\end{alignat}
It matters a lot if the detuning is small or large. This is because
\begin{alignat}{1}
\frac{\partial\alpha}{\partial x}=-\frac{k_L \frac{\delta_L}{\Omega} \sin[k_L(x-\Delta L \sin\omega t)]}{\left( \frac{\delta_L}{\Omega}\right)^2 + \cos^{2}[k_L(x-\Delta L \sin\omega t)]}; \qquad 
\frac{\partial\alpha}{\partial t}=\frac{\omega\frac{\delta_L}{\Omega} \sin[k_L(x-\Delta L \sin\omega t)]\cos\omega t}{\left( \frac{\delta_L}{\Omega}\right)^2 + \cos^{2}[k_L(x-\Delta L \sin\omega t)]}.
\end{alignat}
So either for small or large detuning,
\begin{equation}
\delta_L\ll\Omega \quad {\rm or} \quad \delta_L\gg\Omega \quad \Rightarrow \quad \frac{\partial\alpha}{\partial t},\frac{\partial\alpha}{\partial x}\rightarrow 0.
\label{det_beh}
\end{equation}
\subsubsection{Small detuning}
Here, $\omega _0 \sim \omega _L$, thus $\tan \alpha \to \infty$ or $\alpha \sim \pi/2$. Considering \eqref{det_beh} and keeping the terms upto O$(\hbar)$, the adiabatic Hamiltonian further simplifies to
\begin{alignat}{1}
H_{\rm ad,s}^{\rm RWA} = \frac{p^2}{2m}{\bf I} + \hbar \Omega _{\rm eff} \sigma_z.
\end{alignat}
Exploiting the smallness of detuning, we may expand binomially to obtain
\begin{alignat}{1}\label{large_det_eq}
H_{\rm ad,s}^{\rm RWA,\pm} &= \frac{p^2}{2m} \pm \frac{\hbar \Omega}{2} \cos[k_L(x - \Delta L \sin \omega t)] \left[ 1 + \frac{(\omega _0 - \omega _L)^2}{2\Omega^2 \cos ^2[k_L(x - \Delta L \sin \omega t)]} \right] \nonumber \\ &+ {\mathcal O}\left( \left( \frac{\omega_0 - \omega_L}{\Omega} \right)^3 \right).
\end{alignat}
These provide the two potential energy surfaces on which the two-level system evolves, connected by tunneling. This can be seen by the fact that the intersection of the two curves occurs when $\Omega_{\rm eff}$ is zero, leading to 
\begin{alignat}{1}
x &= \Delta L\,\sin \omega t + \frac{\pi}{2k_L} + i\log \left( \sqrt{1  - \frac{\delta_L^2}{2\Omega^2}} - \frac{\delta_L}{\sqrt{2}\Omega}   \right)\nonumber \\ 
&\simeq \Delta L\,\sin \omega t + \frac{\pi}{2k_L} - i \frac{\sqrt{2}\delta_L}{2\Omega}
\end{alignat}
for small detuning. The binding part of the potential in \eqref{large_det_eq} supports eigenvalues. However, since the Hamiltonian is periodic in time, the eigenvalues are quasienergies. Owing to the imaginary part, these are more precisely ``quasienergy resonances".

\subsubsection{Large detuning}

We consider the case where we have RWA and adiabatic approximation but $\delta_L\gg\Omega$. Then we have the Hamiltonian,
\begin{alignat}{1}
H_{\rm ad,l}^{\rm RWA} &=\begin{pmatrix}
 	p^{2}/2m + \hbar\Omega _{\rm eff} & 0 \\
	0 & p^{2}/2m - \hbar\Omega _{\rm eff} \\
\end{pmatrix}.
\end{alignat}
This can be decomposed into two Hamiltonians:
\begin{alignat}{1}
H_{\rm ad,l}^{\rm RWA,\pm} &=\frac{p^2}{2m} \pm \frac{\hbar\delta_L}{2}\left[ 1 + \frac{\Omega^{2}}{2\delta_{L}^{2}}\cos^2[k_L(x-\Delta L\sin\omega t)]\right]
+ {\mathcal O}\left( \left( \frac{\Omega}{\omega_0 - \omega_L} \right)^3 \right).
\label{no_binom}
\end{alignat}
The potential energy curves intersect when 
\begin{equation}
x(t) = \left( n + \frac{1}{2} \right)\frac{\pi}{k_L} + \Delta L\,\sin \omega t.
\end{equation}
Here the intersection points are real where the real part is the same as for small detuning. The potential energy curves support sharp quasienergies. 

\section{Exact solution}

We now return to the \eqref{eq:6} and lift all the approximations considered in the last Section. The Hamiltonian is written as 
\begin{alignat}{1}\label{eq:hrot}
H_{\rm rot} = \frac{p^2}{2m}{\bf I} + \begin{pmatrix}
 	a & b \\
	b* & -a \\ 
\end{pmatrix} \equiv \frac{p^2}{2m}{\bf I} + \mathcal{M}
\end{alignat}
where $a = \hbar (\omega_0 - \omega_L)/2$, $b=b_1 + ib_2$ with
\begin{alignat}{1}
b_1 &= \frac{\hbar \Omega}{2} \cos[k_L(x - \Delta L \sin \omega t)](1 + \cos 2\omega _Lt), \nonumber \\
b_2 &= \frac{\hbar \Omega}{2} \cos[k_L(x - \Delta L \sin \omega t)]\sin 2\omega _Lt.
\end{alignat}
The matrix, denoted by ${\mathcal M}$ in \eqref{eq:hrot} can be diagonalized by a matrix ${\mathcal S}$ to get the diagonal matrix, ${\mathcal J}$. The matrices are

\begin{alignat}{1}
{\mathcal S} &= \begin{pmatrix}
 	\frac{(a - \sqrt{a^2 + b_1^2 + b_2^2})(b_1 + ib_2)}{b_1^2 + b_2^2} & \frac{(a + \sqrt{a^2 + b_1^2 + b_2^2})(b_1 + ib_2)}{b_1^2 + b_2^2} \\
	1 & 1 \\
\end{pmatrix}
\end{alignat}
and
\begin{alignat}{1}
{\mathcal J} &= \begin{pmatrix}
 	- \sqrt{a^2 + b_1^2 + b_2^2} & 0 \\
	0 & \sqrt{a^2 + b_1^2 + b_2^2} \\
\end{pmatrix}.
\end{alignat}
Define $\psi _1 = {\mathcal S}^{-1} \psi_{\rm rot}$ with $i\hbar \partial \psi_{\rm rot}/\partial t = {\mathcal H}\psi_{\rm rot}$. The equation for the time evolution of $\psi _1$ is 
\begin{alignat}{1}
i\hbar \frac{\partial \psi_1}{\partial t} = -i\hbar {\mathcal S}^{-1}\frac{\partial {\mathcal S}}{\partial t}\psi_1 + {\mathcal S}^{-1}\frac{p^2}{2m}{\bf I}{\mathcal S}\psi_1 + {\mathcal J}\psi_1.
\end{alignat}
Now, ${\mathcal S}^{-1}p^2{\mathcal S} = ({\mathcal S}^{-1}p{\mathcal S})^2 = (p - i\hbar {\mathcal S}^{-1}\partial {\mathcal S}/\partial x)^2$. Here we again have a vector potential which is an artificial gauge field. 

The Hamiltonian is thus written as an expansion \cite{gba,jain2004},
\begin{equation}
H = H_0 + \hbar H_1 + \hbar^2 H_2  
\label{exact}  
\end{equation}
with $H_0$ has a simple form:
\begin{alignat}{1}
H_0 &= \frac{p^2}{2m}\textbf{I} + \begin{pmatrix}
 	- \sqrt{a^2 + b_1^2 + b_2^2} & 0 \\
	0 & \sqrt{a^2 + b_1^2 + b_2^2}. \\
\end{pmatrix}
\end{alignat}
Writing $\psi_1 = (\psi_1^{(+)} ~~~\psi_1^{(-)})^{T}$ with the superscript, $T$ denoting the transpose, we have written the state with two components. The classical Hamiltonians corresponding to the states, $\psi_1^{(\pm)}$ are
\begin{alignat}{1}\label{eq:60}
H_0^{(\pm)} = \frac{p^2}{2m} \pm \frac{\hbar(\omega_0 - \omega_L)}{2} \left( 1 + \frac{4\Omega^2}{(\omega_0 - \omega_L)^2}\cos ^2[k_L(x - \Delta L \sin \omega t)]\cos ^2 \omega_Lt \right)^{1/2}.
\end{alignat}
Usually, $\psi_1^{(+)}$ is subjected to a binding potential and $\psi_1^{(-)}$ is evolving on a scattering potential. There are two potential energy surfaces, $\pm \sqrt{a^2 + b_1^2 + b_2^2}$ 
on which the full two-component wavefunction, $\psi_1$ evolves. The potential energy surfaces meet at the solution of 
\begin{equation}
a^2 + b_1^2 + b_2^2 = 0.
\end{equation}
\begin{figure}[h]
\includegraphics[width=\textwidth]{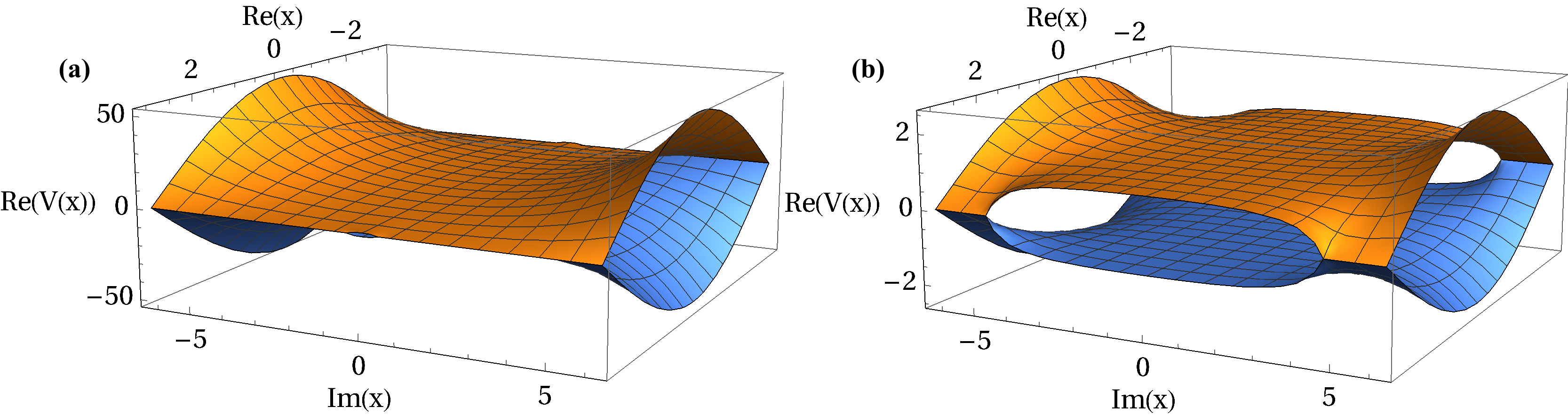}
\caption{Potential Energy Surface (PES) at (a) large detuning ($\delta_{L}\gg \Omega$) and (b) small detuning ($\delta_{L}\ll \Omega$). At large detuning the gap shrinks allowing a larger region for space for crossing of PES}\label{PES_lsd}
\end{figure}
The solution is
\begin{alignat}{1}
x &= \Delta L \sin \omega t + \frac{1}{k_L} \cos^{-1} \left[ \pm i\, \frac{(\omega_0 - \omega_L)}{2\Omega} \sec (\omega_Lt) \right] \nonumber \\
&= \Delta L \sin \omega t + \frac{\pi}{2 k_L} + i\,\frac{1}{k_L}\log \left[ 1 \mp \frac{\delta_L}{2\Omega}\sec (\omega_Lt) + \frac{\delta_L^2}{8\Omega^2}\sec ^2(\omega_L t)\right].
\end{alignat}
For small detuning ($\delta_L \ll \Omega$), the potential curves intersect at
\begin{equation}
x = \Delta L \sin \omega t + \frac{\pi}{2k_L} \mp i\,\frac{\delta_L}{2\Omega}\sec (\omega_Lt) \pm i\,\frac{\delta_L^3}{48\Omega^3}\sec^3 (\omega_Lt).
\end{equation}

\begin{figure*}[ht]
\centering
\includegraphics[width=1.0\textwidth]{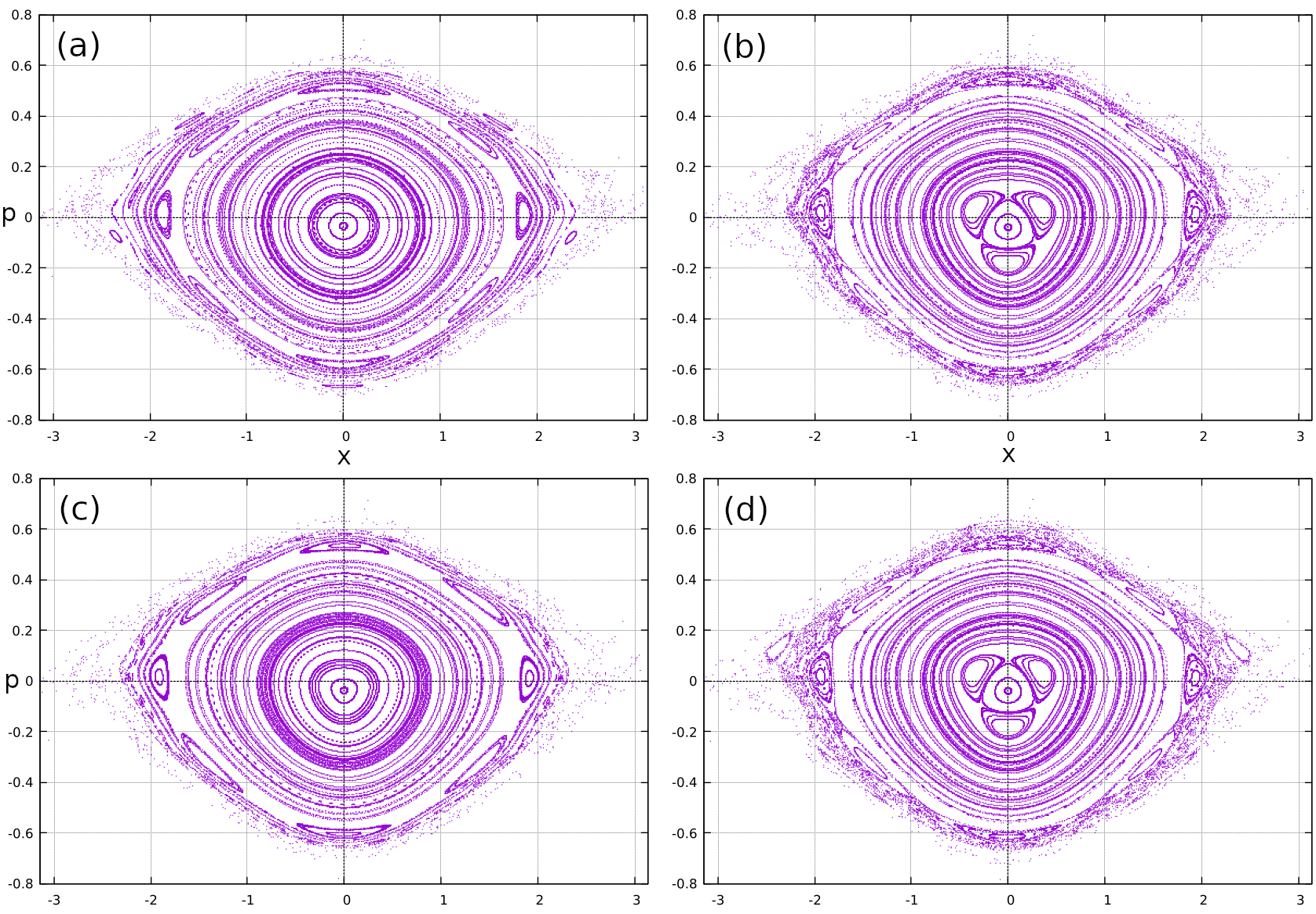}
\caption{Comparison of Poincare sections for Hamiltonians under different approximations for the case of large detuning for the same set of parameters used in Fig.~\ref{poincare}. (a) Shows the un-approximated case corresponding to the exact solution. (b) Shows the application of binomial approximation to the exact solution. (c) Corresponds to the RWA+Adiabatic approximation and (d) corresponds to the RWA+adiabatic+Binomial approximation. Initial conditions and number of evolution steps are kept the same for all cases here.}
\label{comp}
\end{figure*}
The complex value of crossing of the potential energy surfaces implies the tunneling of atoms. The tunneling across these surfaces where the underlying dynamics is nonlinear has some very interesting related phenomena like resonance assisted tunneling \cite{BRODIER200288}, which have been recently experimentally realized \cite{PhysRevLett.115.104101}.

The Fig.~\ref{PES_lsd} (a) and (b) show these crossings along the complex position plane. We note that the crossing gap at the null imaginary position plane vanishes as one reaches closer to resonance (at small detuning) and remains wide open at large detuning.

In \eqref{eq:60}, for large detuning, $\Omega^2/(\omega_0-\omega_L)^2 \ll 1$, a Taylor expansion immediately yields 
\begin{alignat}{1}\label{eq:66}
H_{0,l}^{(\pm)} = \frac{p^2}{2m} \pm \frac{\hbar(\omega_0 - \omega_L)}{2} \left( 1 + \frac{2\Omega^2}{(\omega_0 - \omega_L)^2}\cos ^2[k_L(x - \Delta L \sin \omega t)]\cos ^2 \omega_Lt \right).
\end{alignat}
Among the two Hamiltonians, $H_{0,l}^{(-)}$ is binding; it can be seen that the second term in the Taylor expansion of $\cos [k_L(x - \Delta L \sin \omega t)]$ along with an overall negative sign will make this roughly parabolic for small arguments, at least. For the same reason, $H_0^{(+)}$ is a scattering potential. The differences in Poincare sections for various cases can be seen in the following figure. We found that the 3 island ring which is present in both un-approximated case and RWA+Adiabatic case vanishes if we make a binomial approximation implying origin of this resonance is purely because of higher order terms of \eqref{eq:66} and \eqref{large_det_eq}. We also note that the chaos is more apparent in the binomial case but less severe in all other cases.

We now study the classical mechanics of these Hamiltonians. 

\section{Quasi-integrability}

In this Section, we study the classical dynamics of the  Hamiltonians obtained above under different approximations.

We begin with the exact Hamiltonian, namely \eqref{exact}, and consider only $H_0^{(-)}$ in \eqref{eq:60}. We make the following transformations to convert it to a dimensionless form almost similar to \cite{stockmann_1999}.
\begin{alignat}{1}
&t\rightarrow \frac{t}{\omega}~,~x\rightarrow\frac{x}{2k_{L}}~,~p\rightarrow\frac{M\omega p}{2k_{L}}~,~H_{0}^{-}\rightarrow \frac{M \omega^{2} H_{0}^{-}}{4 K_{L}^{2}} \nonumber \\
&\lambda=2k_{L}\Delta L ~,~ \gamma=\frac{\omega_{L}}{\omega}~,~\eta=\left( \frac{\Omega}{\delta_{L}}\right)^{2}~,~K=\frac{\hbar k_{L}^{2}\Omega^{2}}{2M\omega^{2}\delta_{L}}
\label{transformers}
\end{alignat}
where $\eta$ is strength of Rabi resonance and $\delta_{L}=\omega_{0}-\omega_{L}$ is the detuning of laser. The simplified Hamiltonian yields:
\begin{equation}
H_{0}^{-}=\frac{p^{2}}{2}-\frac{4K}{\eta}\left[1+2\eta(1+\cos(x-\lambda\sin t))\cos^2 \gamma t \right]^\frac{1}{2}
\end{equation}
Now, using the same transformations \eqref{transformers}, we write the  Hamiltonians for large detuning, neglecting the constant terms:
\begin{alignat}{1}
&H_{0,l}^{-}\simeq\frac{p^{2}}{2}-4K\cos(x-\lambda\sin t)\cos^2 \gamma t, \\
&H_{\rm ad,l}^{\rm RWA,-}\simeq\frac{p^{2}}{2}-K\cos(x-\lambda\sin t).
\label{dim_less_large}
\end{alignat}
This clearly implies a drastic change in the equation since if $\gamma\gg 1$, thus even if we use $\langle \cos^2\gamma t\rangle=1/2$, the second term contributes double compared to the contribution coming from the usual case with adiabatic and RWA approximation.

\begin{figure}[h]
\centering
\includegraphics[width=\textwidth]{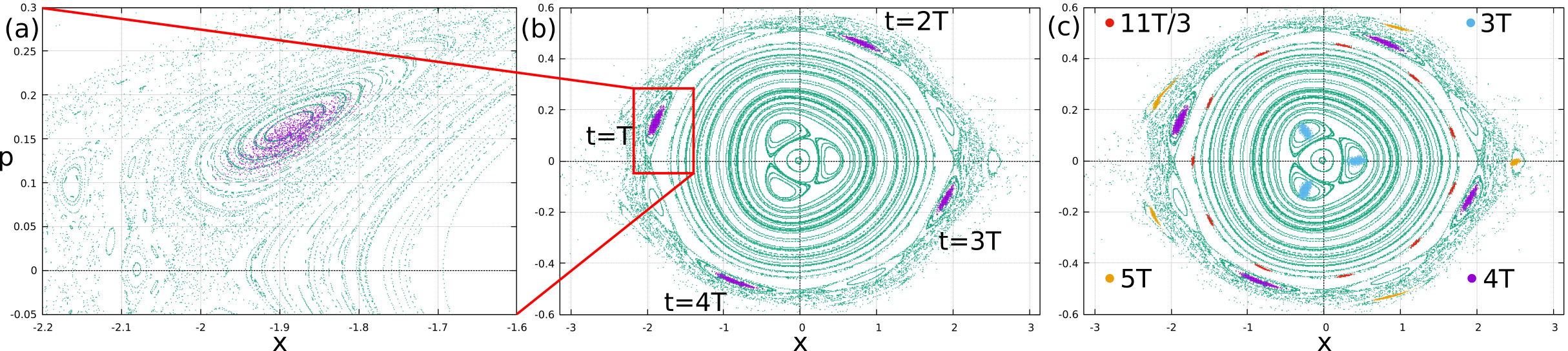}
\caption{Poincar\'{e} Sections taken in steps of modulation period using the same parameter as in \cite{PhysRevLett.73.2974}. (a) 1000 ultracold atoms (purple dots) are loaded in one of the islands of stability in the Poincare section taken in steps of the driving period T. (b) stroboscopic evolution of the ultracold atoms reveals that they evolve with a period 4T. (c) Similarly, loading on different islands of stability shows the existence of 3T, 11T/3, 4T and 5T periods predominantly.}
\label{poincare}
\end{figure}

In order to understand the underlying phase space structure, we initialize 1000 ultracold atoms (blue dots) in one of the island in the Poincar\`{e} section taken in steps of modulation time period $T$ as shown in Fig.~\ref{poincare} (top) and look at its stroboscopic evolution in multiples of the modulation time period. We find that after each modulation period, atoms move from one island to another lying around the same larger elliptic-like orbit (Fig.~\ref{poincare} (middle)). Similarly, we find that the number of islands is equal to (or twice if $n$ is even) the number of modulation periods $n$ for the marked islands in Fig.~\ref{poincare} (bottom). In other words, these islands satisfies $T_{\rm orbit}=nT$ or $\Omega_{\rm orbit}/\omega =1/n$.

To study the origin of these patterns in resonance structures, we write the dimensionless Hamiltonian \eqref{dim_less_large} in action-angle variables. Let us write one of the RWA Hamiltonians as a perturbed harmonic oscillator: 
\begin{alignat}{1}
H^{\rm RWA,-}_{\rm 0,l} &=\frac{p^2}{2} + \frac{K x^{2}}{2} - \left( K\cos(x-\lambda\sin t) + \frac{K x^{2}}{2} \right)\\
&=H_{\rm h.o.} + \epsilon\Delta H.
\end{alignat}
where $\epsilon$ is introduced for book-keeping (eventually, we shall put $\epsilon=1$). Employing the oscillator  action-angle variables, $(J, \theta)$, with  $x=\sqrt{\frac{J}{\pi\Omega}}\sin(\theta)$ and $p=\sqrt{\frac{J\Omega}{\pi}}\cos(\theta)$ with $K=\Omega^2$, the Hamiltonians are:
\begin{alignat}{1}
H_{h.o.}&=\frac{\Omega J}{2\pi}~~\\
\Delta H &= -\Omega^{2}\cos\left(\sqrt{\frac{J}{\pi\Omega}}\sin\theta - \lambda\sin t\right) 
-\frac{J\Omega}{2\pi}\sin^{2}\theta .
\end{alignat}

We use the classical time-dependent perturbation theory \cite{Lichtenberg1992} to calculate the associated action of this Hamiltonian up to first order in perturbation. For this, we transform the action variables in a way that the new Hamiltonian $\bar{H}$ is only a function of the new action variable $\bar{J}$ alone. We obtain
\begin{alignat}{2}
\langle\Delta H\rangle &= \frac{1}{2\pi}\int_{0}^{2\pi}dt\frac{1}{2\pi}\int_{0}^{2\pi}d\theta\Delta H (J,\theta, t)\nonumber\\
&= -\Omega^{2}J_{0}\left(\sqrt{\frac{\bar{J}}{\Omega\pi}}\right)J_{0}(\lambda) - \frac{\bar{J}\Omega}{4\pi}\\
\bar{H}(\bar{J})&= \frac{\Omega\bar{J}}{2\pi} - \epsilon\Omega^{2}J_{0}\left(\sqrt{\frac{\bar{J}}{\Omega\pi}}\right)J_{0}(\lambda) - \epsilon\frac{\bar{J}\Omega}{4\pi}
\end{alignat}
where $J_0(.)$ is the cylindrical Bessel function of order zero. 
The new frequency is
\begin{equation}
 {\Omega}'(\bar{J})=2\pi\frac{\partial\bar{H}}{\partial\bar{J}}=\Omega(1-\epsilon/2) - 2\epsilon\pi\Omega^{2}J_{0}'\left(\sqrt{\frac{\bar{J}}{\Omega\pi}}\right)J_{0}(\lambda) \end{equation}
where prime on the Bessel function denotes a derivative with respect to its argument. 

\begin{figure}[h]
\centering
\includegraphics[width=0.50\textwidth]{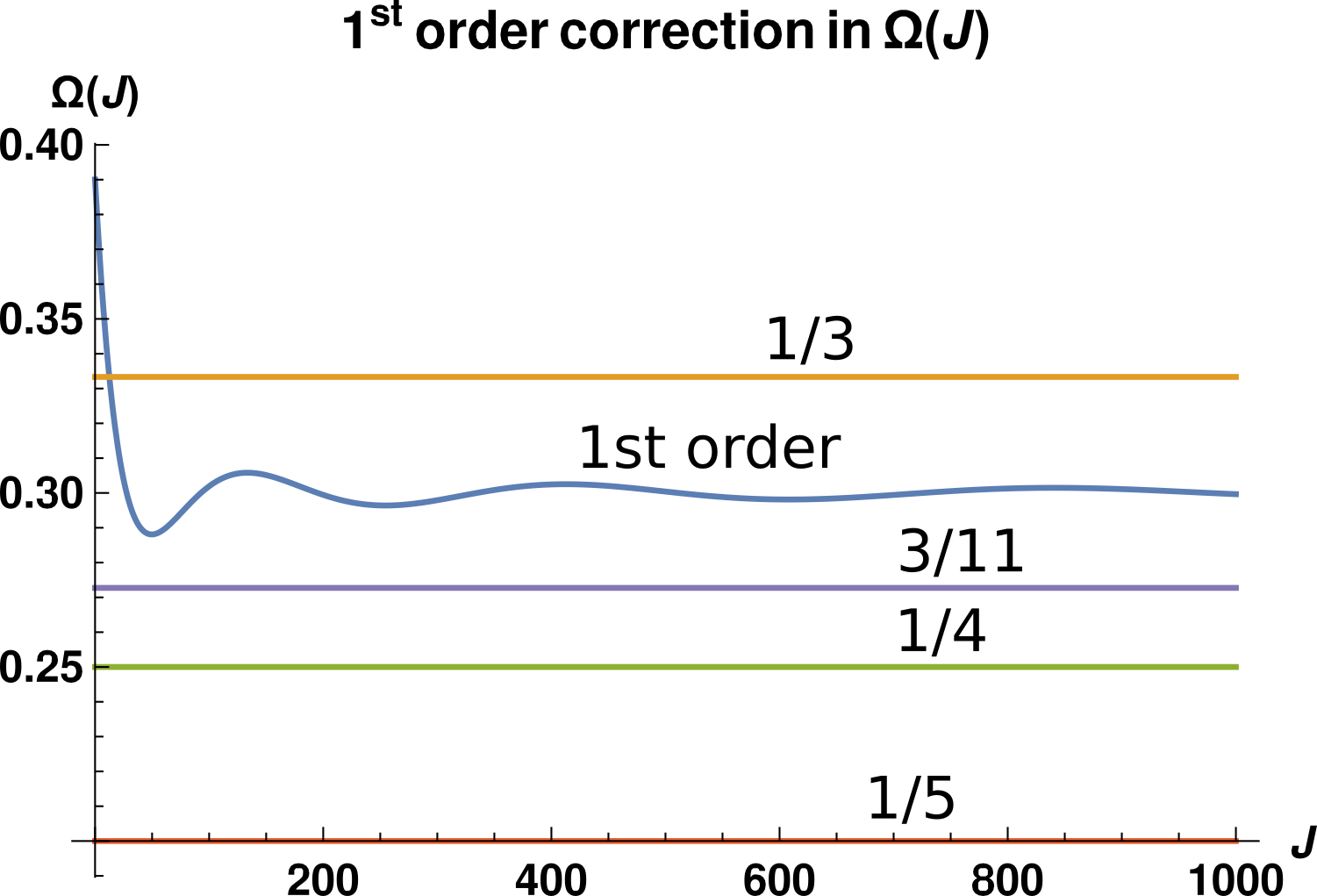}
\caption{Only those resonances whose frequency ratio $\Omega(J) : \omega$ ($\omega$=1 here) intersect with the $\Omega(J)$ are allowed.}
\label{omegaj}
\end{figure}

We subtract this $\epsilon\langle\Delta H\rangle$ from $\epsilon\Delta H$ to obtain the oscillating part $\epsilon\lbrace\Delta H \rbrace$. For calculating the integral, we expand the potential term using Jacobi-Anger expansion \cite{abramowitz}  $e^{iz\sin\theta}=\sum_{n=-\infty}^{+\infty} J_{n}(z)e^{i n\theta}$:

\begin{alignat}{2}
\lbrace \Delta H \rbrace &= -\sum_{n,m=-\infty}^{\infty}\Omega^{2}J_{n}\left(\sqrt{\frac{\bar{J}}{\Omega\pi}}\right)J_{m}(\lambda)\cos(n\bar{\theta}-m t)
+ \frac{\bar{J}\Omega}{4\pi}\cos 2\bar{\theta}\\
 &\equiv  \sum_{n,m=-\infty}^{\infty} \Delta H_{n,m}(\bar{J},\bar{\theta},t) + \frac{\bar{J}\Omega}{4\pi}\cos 2\bar{\theta}
\end{alignat}
where both $n, m$ are non-zero. The change in action $\epsilon\Delta S$ can be calculated as
\begin{alignat}{2}
\epsilon\Delta S &=-\int^{t} dt \epsilon\lbrace \Delta H \rbrace\\
&=\sum_{n,m=-\infty}^{\infty}\epsilon\Delta S_{n,m}(\bar{J},\bar{\theta},t) + \frac{\epsilon\bar{J}\Omega}{8\pi\bar{\Omega}(\bar{J})}\sin 2\bar{\theta}
\end{alignat}
where
\begin{alignat}{1}
\epsilon\Delta S&_{n,m}=\frac{-\epsilon\Omega^{2}}{n\bar{\Omega}(\bar{J})-m}J_{n}\left(\sqrt{\frac{\bar{J}}{\Omega\pi}}\right)J_{m}(\lambda)\sin(n\bar{\theta}-m t)
\end{alignat}
Consequent to the above,
\begin{equation}
\bar{J}=J-\epsilon\frac{\partial \Delta S}{\partial \theta}(J,\theta,t)~;~\bar{\theta}=\theta + \epsilon\frac{\partial \Delta S}{\partial J}(J,\theta,t).
\end{equation}
The new action-angle variables can be calculated up to first order as
\begin{alignat}{2}
\bar{J}&=J + \epsilon\frac{n\Omega^{2}}{n\bar{\Omega}(J)-m}J_{n}\left(\sqrt{\frac{J}{\Omega\pi}}\right)J_{m}(\lambda)\cos(n{\theta}-m t) 
-\epsilon\frac{J\Omega}{4\pi}\cos2\theta , \\
\bar{\theta}&=\theta + \epsilon\frac{-\Omega^{2}}{n\bar{\Omega}(J)-m}J_{n}'\left(\sqrt{\frac{J}{\Omega\pi}}\right)J_{m}(\lambda)\sin(n\theta-m t) + \frac{\epsilon\Omega}{8\pi\bar{\Omega}(\bar{J})}\sin 2\theta .
\end{alignat}
Thus we have obtained the action with resonant denominators which leads to resonant condition
\begin{equation}
n\bar{\Omega}(\bar{J})=m\omega
\end{equation}
where $\omega$ is the modulation frequency and $\bar{\Omega}(\bar{J})$ is the frequency of the orbit, $\omega$ is obtained when we substitute actual time, $t$ in place of dimensionless time from \eqref{transformers}. This explains the observed pattern in Fig.~\ref{poincare} : the orbital periods are integral multiples of the modulation period at the resonance. The strength of $\rm (n,m)^{\rm th}$ resonance is determined by the product of two Bessel functions $J_{n}(\sqrt{J/\Omega\pi})$ and $J_{m}(\lambda)$. Using the first-order correction in the frequency $\Omega(J)$, we plot it as a function of $J$ in Fig.~\ref{omegaj}. We see that only the 1:3 resonance is allowed under first-order correction. This means that all other resonances in Fig.~\ref{poincare} must originate from the higher-order perturbation terms in correction for $\bar{\Omega}$ and $\bar{J}$. That explains the dominance of primary islands in (n,m)=(3,1) resonance and the presence of secondary islands in other resonances.

For the expression without binomial approximation \eqref{dim_less_large}  where in Fig.\ref{comp} we saw (3,1) resonance to be dominantly present, but without binomial approximation \eqref{no_binom}, this resonance is suppressed and doesn't appear. This can lead to significant corrections for both quantum and classical equations despite being in large detuning limit. Similarly, very high-ordered resonances are enhanced by binomial approximation as the chaotic regime can be seen enhanced around the edges for this case.

\section{Dynamical localization}

Let us imagine that we prepare the initial state of the atoms as a localized wavepacket. As the system evolves, the wavepacket spreads. The wavefunction of the two-state system is shown to evolve, in all versions of description, on a pair of potential energy surfaces. The  form of these potentials readily support bounded dynamics on one of the potentials. The complex intersection points provide paths for tunneling. The succession of these two dynamical features leads to localization of the wavepacket. The physics of this is nothing but the well-known argument by Mott \cite{mott} and Anderson \cite{anderson1958absence}, adapted in recent times in quantum chaos \cite{fgp,jain1993}.   

\section{Conclusions}

The matrix Hamiltonian driving a two-level atom has been unitarily transformed to a series of Hamiltonians arranged in the powers of Planck constant - which is the precise meaning of semiclassical expansion. A successive application of these transformations brings out an effective Hamiltonian to any desired level of accuracy. In principle, one could perform computations to all orders of $\hbar$. The system is shown to tunnel between two potential energy surfaces, the underlying dynamics is quasi-integrable in the KAM sense. 

The analysis has been carried out in the past by employing physically appealing and rather standard approximations. We recapitulated these and then have provided exact solution where by ``exact", we mean in the sense described in the preceding paragraph. We have seen that a matrix Hamiltonian for a spinor eigenstate. At different orders of Planck constant, there are different potential energy surfaces on which the system is shown to evolve. If one makes a binomial approximation in the Hamiltonian to treat the system, the detailed features in the Poincar\'{e} surfaces of section differ. The approximated analysis has certain appeal insofar as tunneling between islands is seen clearly. However, to establish that the existence of islands and tunneling, We show that the onset of islands of stability can be seen from the first-order perturbation theory. 

The analysis reveals a vector potential that is related to an artificial gauge field. We believe that knowing the form for this could be useful for experiments with cold atoms and in developing fields of Hamiltonian engineering, quantum sensing and quantum interference. We have not developed these aspects here. 

As referred to in the Introduction, our results add to the discussion of integrability in matrix models for atomic systems, in particular to the work on quantum Rabi model \cite{braak}. In the future, by adding nonlinear terms to incorporate interactions that allow control of atomic states, these works could be useful for critical quantum metrology \cite{braak_nl}. Control of states of multi-qubit systems \cite{krjj} and their protection \cite{ssj} belongs to the present theme in a rather compelling manner.   

\vskip 1.0 truecm
\noindent
{\bf Acknowledgements}
We thank Sandeep Joshi for several helpful discussions. RG acknowledges the fellowship support received from CSIR-HRDG.


\begin{thebibliography}{99}

\bibitem{PhysRevLett.73.2974} Moore F L, Robinson J C, Bharucha C, Williams P E, and Raizen M G. 1994 Observation of Dynamical Localization in Atomic Momentum Transfer: A New Testing Ground for Quantum Chaos. \textit{Phys. Rev. Lett.} \textbf{73}, 2974. 

\bibitem{thalhammer2008double} Thalhammer G, Barontini G, De Sarlo L, Catani J, Minardi F, and Inguscio M. 2008 Double species Bose-Einstein condensate with tunable interspecies interactions. \textit{Phys. Rev. Lett. } \textbf{100}, 210402. 

\bibitem{cui2018broad} Yue Cui, Deng Min, You Li, Gao Bo and Tey Meng Khoon 2018 Broad Feshbach resonances in ultracold alkali-metal systems. \textit{Phys. Rev. A} \textbf{98}, 042708.

\bibitem{GiovPhD} Barontini G. 2010 \textit{PhD thesis}, \textit{University of Florence}

\bibitem{tanzi2018feshbach} Tanzi L, Cabrera C R, Sanz J, Cheiney P, Tomza M and Tarruell L 2018 Feshbach resonances in potassium Bose-Bose mixtures. \textit{Phys. Rev. A} \textbf{98}, 062712.

\bibitem{d2007feshbach} D'Errico C, Zaccanti M, Fattori M, Roati G, Inguscio M, Modugno G and Simoni A 2007 Feshbach resonances in ultracold $^{39}$K. \textit{New Journal of Physics} \textbf{9}, 223.

\bibitem{marte2002feshbach} Marte A, Volz T, Schuster J, D{\"u}rr S, Rempe G, Van Kempen E G M and Verhaar B J 2002 Feshbach Resonances in Rubidium 87: Precision Measurement and Analysis. \textit{Phys. Rev. Lett.} \textbf{89}, 283202.

\bibitem{yurovsky2003three} Yurovsky V A and Ben-Reuven A 2003 Three-body loss of trapped ultracold $^{87}$Rb atoms due to a Feshbach resonance. \textit{Phys. Rev. A} \textbf{67}, 050701.

\bibitem{vogels1997prediction} Vogels J M, Tsai C, Freeland R, Kokkelmans S, and Verhaar B and Heinzen D 1997 Prediction of Feshbach resonances in collisions of ultracold rubidium atoms. \textit{Phys. Rev. A} \textbf{56}, R1067. 

\bibitem{blackley2013feshbach} Blackley C L, Le Sueur C R, Hutson J M, McCarron D J, K{\"o}ppinger M P, Cho Hung-Wen, Jenkin D L and Cornish S L 2013 Feshbach resonances in ultracold $^{85}$Rb. \textit{Phys. Rev. A} \textbf{87}, 033611.

\bibitem{franzen2022observation} Franzen Tobias, Guttridge Alexander, Wilson Kali E, Segal Jack, Frye Matthew D, Hutson Jeremy M and Cornish Simon L 2022 Observation of magnetic Feshbach resonances between Cs and $^{173}$Yb. \textit{Physical Review Research} \textbf{4}, 043072.

\bibitem{cho2013feshbach} Cho Hung-Wen, McCarron Daniel J, K{\"o}ppinger Michael P, Jenkin Daniel L, Butler Kirsteen L, Julienne Paul S, Blackley Caroline L, Le Sueur C Ruth, Hutson Jeremy M and Cornish Simon L 2013 Feshbach spectroscopy of an ultracold mixture of $^{85}$Rb and $^{133}$Cs. \textit{Phys. Rev. A} \textbf{87}, 010703.

\bibitem{catani2008degenerate} Catani J, De Sarlo L, Barontini G, Minardi F and Inguscio M 2008 Degenerate Bose-Bose mixture in a three-dimensional optical lattice. \textit{Phys. Rev. A} \textbf{77}, 011603.

\bibitem{wacker2016universal} Wacker L, J{\o}rgensen N, Birkmose D, Winter N, Mikkelsen M, Sherson J, Zinner N and Arlt J J 2016 Universal three-body physics in ultracold KRb mixtures. \textit{Phys. Rev. Lett.} \textbf{117}, 163201.

\bibitem{ferrari2002collisional} Ferrari G, Inguscio M, Jastrzebski W, Modugno G, Roati G and Simoni A 2002 Collisional properties of ultracold K-Rb mixtures. \textit{Phys. Rev. Lett.} \textbf{89}, 053202.

\bibitem{sawant2021thermalization} Sawant Rahul, Maffei Anna and Barontini Giovanni A 2021 Thermalization of a trapped single atom with an atomic thermal bath. \textit{Appl. Sci.} \textbf{11}, 2258.

\bibitem{mosk2001mixture} Mosk A, Kraft S, Mudrich M, Singer K, Wohlleben W, Grimm R and Weidem{\"u}ller M 2001 Mixture of ultracold lithium and cesium atoms in an optical dipole trap. \textit{Appl. Phy. B} \textbf{73}, 791.

\bibitem{kasevich1991atomic} Kasevich Mark and Chu Steven 1991 Atomic interferometry using stimulated Raman transitions. \textit{Phys. Rev. Lett.} \textbf{67}, 181.

\bibitem{rosi2018detecting} Rosi G, Vicer{\'e} A, Cacciapuoti L, D’Amico G, Hu L, Jain M, Poli N, Salvi L, Sorrentino F, Wang E and Tino G M 2018 Detecting gravitational waves with atomic sensors. \textit{Nuovo Cimento della Societ{\`a} Italiana di Fisica C, Geophysics and Space Physics} \textbf{41}, 130.

\bibitem{rosi2017proposed} Rosi G 2017 A proposed atom interferometry determination of G at 10$^{-5}$ using a cold atomic fountain. \textit{Metrologia} \textbf{55}, 50.

\bibitem{rosi2014precision} Rosi G, Sorrentino F, Cacciapuoti L, Prevedelli M and Tino G M 2014 Precision measurement of the Newtonian gravitational constant using cold atoms. \textit{Nature} \textbf{510}, 518.

\bibitem{rosi2017quantum} Rosi G, D’Amico G, Cacciapuoti L, Sorrentino F,  Prevedelli M, Zych M, Brukner {\v{C}} and Tino G M 2017 Quantum test of the equivalence principle for atoms in coherent superposition of internal energy states. \textit{Nature communications} \textbf{8}, 15529.

\bibitem{jain2022reply} Jain M, Tino G M, Cacciapuoti L and Rosi G 2022 Reply to comment on “New apparatus design for high precision measurement of G with atom interferometry”. \textit{The European Physical Journal D} \textbf{76}, 164.

\bibitem{jain2021new} Jain M, Tino G M, Cacciapuoti L and Rosi G 2021 New apparatus design for high-precision measurement of G with atom interferometry. \textit{The European Physical Journal D} \textbf{75}, 197.

\bibitem{d2019measuring} D’Amico Giulio, Cacciapuoti Luigi, Jain Manan, Zhan Su and Rosi Gabriele 2019 Measuring the gravitational acceleration with precision matter-wave velocimetry. \textit{The European Physical Journal D} \textbf{73}, 98.

\bibitem{salvi2018testing} Salvi Leonardo, Cacciapuoti Luigi, D'Amico Giulio, Hu Liang, Jain Manan, Poli Nicola, Rosi Gabriele, Wang Enlong and Tino Guglielmo M 2018 Testing gravity with atomic quantum sensors on ground and in space. \textit{Quantum Technologies} \textbf{10674}, 14.

\bibitem{tino2021testing} Tino Guglielmo M 2021 Testing gravity with cold atom interferometry: results and prospects. \textit{Quantum Science and Technology} \textbf{6}, 024014.

\bibitem{lamporesi2008determination} Lamporesi G, Bertoldi A, Cacciapuoti L, Prevedelli M and Tino Guglielmo M 2008 Determination of the Newtonian gravitational constant using atom interferometry. \textit{Phys. Rev. Lett.} \textbf{100}, 050801. 

\bibitem{liu2019vortex} Liu Jixun, Wang Xi, Mellado Mu{\~n}oz Jorge, Kowalczyk Anna and Barontini Giovanni 2019  Vortex conveyor belt for matter-wave coherent splitting and interferometry. \textit{Scientific Reports} \textbf{9}, 1267.

\bibitem{canuel2006six} Canuel B, Leduc F, Holleville D, Gauguet A, Fils J, Virdis A, Clairon A, Dimarcq N, Bord{\'e} Ch J and Landragin A 2006 Six-axis inertial sensor using cold-atom interferometry. \textit{Phys. Rev. Lett.} \textbf{97}, 010402.

\bibitem{aguilera2014ste} Aguilera D, Ahlers H, Battelier B, Bawamia A, Bertoldi A, Bondarescu R, Bongs K, Bouyer P, Braxmaier C and Cacciapuoti L and others 2014. \textit{Classical and Quantum Gravity} \textbf{31}, 115010. 

\bibitem{biedermann2015testing} Biedermann G, Wu X, Deslauriers L, Roy S, Mahadeswaraswamy C and Kasevich M 2015 Testing gravity with cold-atom interferometers. \textit{Phys. Rev. A} \textbf{91}, 033629.

\bibitem{kale2022field} Kale Yogeshwar B, Singh Alok, Gellesch Markus, Jones Jonathan M, Morris David, Aldous Matthew, Bongs Kai and Singh Yeshpal 2022 Field deployable atomics package for an optical lattice clock. \textit{Quantum Science and Technology} \textbf{7}, 045004.

\bibitem{clockchapterbook} Sun Qiushuo, Jones Jonathan M, Singh Alok, Gellesch Markus, Kale Yogeshwar, Singh Vijay, Barron Richard, Jain Manan, Bongs Kai and Singh Yeshpal 2023. \textit{Frequency \& Time: Measurements, Control and Transfer} \textbf{1}, Chapter 2, IFSA Publishing.

\bibitem{ludlow2015optical} Ludlow Andrew D, Boyd Martin M, Ye Jun, Peik Ekkehard and Schmidt Piet O 2015 Optical atomic clocks. \textit{Reviews of Modern Physics} \textbf{87}, 637. 

\bibitem{barontini2022measuring} Barontini G, Blackburn L, Boyer V, Butuc-Mayer F, Calmet X, L{\'o}pez-Urrutia  J C, Crespo E, Darqui{\'e} B and Dunningham J and Fitch N and others 2022 Measuring the stability of fundamental constants with a network of clocks. \textit{EPJ Quantum Technology} \textbf{9}, 12.

\bibitem{takamoto2005optical} Takamoto Masao, Hong Feng-Lei, Higashi Ryoichi and Katori Hidetoshi 2005 An optical lattice clock. \textit{Nature} \textbf{435}, 321.

\bibitem{borde2002atomic} Bord{\'e} Ch J 2002 5D relativistic atom optics and interferometry. \textit{Metrologia} \textbf{39}, 435.

\bibitem{ushijima2015cryogenic} Ushijima Ichiro, Takamoto Masao, Das Manoj, Ohkubo Takuya and Katori Hidetoshi 2015 Cryogenic optical lattice clocks. \textit{Nature Photonics} \textbf{9}, 185.

\bibitem{derevianko2011colloquium} Derevianko Andrei and Katori Hidetoshi 2011 Colloquium: Physics of optical lattice clocks. \textit{Reviews of Modern Physics} \textbf{83}, 331.

\bibitem{yamanaka2015frequency} Yamanaka Kazuhiro, Ohmae Noriaki, Ushijima Ichiro, Takamoto Masao and Katori Hidetoshi 2015 Frequency Ratio of $^{199}$Hg and $^{87}$Sr Optical Lattice Clocks beyond the SI Limit \textit{Phys. Rev. Lett.} \textbf{114}, 230801.

\bibitem{zheng2022differential} Zheng Xin, Dolde Jonathan, Lochab Varun, Merriman Brett N, Li Haoran and Kolkowitz Shimon 2022 Differential clock comparisons with a multiplexed optical lattice clock. \textit{Nature} \textbf{602}, 425.

\bibitem{poli2013optical} Poli Nicola, Oates C W, Gill Patrick and Tino G M 2013 Optical atomic clocks. \textit{La rivista del Nuovo Cimento} \textbf{36}, 555.

\bibitem{zhang1993quantum} Zhang Weiping and Walls D 1993 Quantum diffraction of ultracold atoms by a standing wave laser. \textit{Quantum Optics: Journal of the European Optical Society Part B} \textbf{5}, 9.

\bibitem{jain2018classical} Jain Manan 2018 Classical system underlying a diffracting quantum billiard. \textit{Pramana} \textbf{90}, 20.

\bibitem{Giov1} Barontini Giovanni and Paternostro Mauro 2019 Ultra-cold single-atom quantum heat engines. \textit{New J. Phy.} \textbf{21}, 063019.

\bibitem{munoz2020dissipative} Mu{\~n}oz Jorge Mellado, Wang Xi, Hewitt Thomas, Kowalczyk Anna, Sawant Rahul and Barontini Giovanni 2020 Dissipative distillation of supercritical quantum gases. \textit{Phys. Rev. Lett.} \textbf{125}, 020403.

\bibitem{glatthard2022optimal} Glatthard Jonas, Rubio Jes{\'u}s,  Sawant Rahul, Hewitt Thomas, Barontini Giovanni and Correa Luis A 2022 Optimal cold atom thermometry using adaptive bayesian strategies. \textit{PRX Quantum} \textbf{3}, 040330.

\bibitem{vogler2013thermodynamics} Vogler Andreas, Labouvie Ralf, Stubenrauch Felix, Barontini Giovanni, Guarrera Vera and Ott Herwig 2013 Thermodynamics of strongly correlated one-dimensional Bose gases. \textit{Phys. Rev. A} \textbf{88}, 031603.

\bibitem{tim} Softley T P 2023 Cold and ultracold molecules in the twenties. \textit{Proc. R. Soc. A} \textbf{479}, 20220806.

\bibitem{vijayG} Singh Vijay 2021 Theoretical investigation of a two-stage buffer gas cooled beam source. \textit{Cryogenics} \textbf{118}, 103335.

\bibitem{ozawa2019topological} Ozawa T and Price H M. 2019 Topological quantum matter in synthetic dimensions. \textit{Nature Reviews Physics} \textbf{1}, 349. 

\bibitem{anderson1958absence} Anderson P W. 1958 Absence of diffusion in certain random lattices. \textit{Phys. Rev.} \textbf{109}, 1492. 

\bibitem{fgp} Fishman S, Grempel D R and Prange R E. 1982 Chaos, quantum recurrences, and Anderson localization. \textit{Phys. Rev. Lett. } \textbf{49}, 509. 

\bibitem{jain1993} Jain S R. 1993 Fractal-like quasienergy spectrum in the Fermi-Ulam model. \textit{Phys. Rev. Lett. } \textbf{70}, 3553. 

\bibitem{billy2008direct} Billy Juliette, Josse Vincent, Zuo Zhanchun, Bernard Alain, Hambrecht Ben, Lugan Pierre, Cl{\'e}ment David, Sanchez-Palencia Laurent, Bouyer Philippe, and Aspect Alain. 2008 Direct observation of Anderson localization of matter waves in a controlled disorder. \textit{Nature} \textbf{453}, 891. 

\bibitem{Lichtenberg1992} Lichtenberg A J, Lieberman M A. 1992 \textit{Regular and Chaotic Dynamics}. Springer Verlag. 

\bibitem{berry_aip} Berry M V. 1978 Regular and irregular motion. \textit{AIP Conference Proceedings} \textbf{46}, 16.

\bibitem{abramowitz} Abramowitz M, Stegun I. 1965 \textit{Handbook of Mathematical Functions with Formulas, Graphs, and Mathematical Tables}. Dover.

\bibitem{braak} Braak D. 2011 Integrability of the Rabi model. \textit{Physical Review Letters} \textbf{107}, 100401.

\bibitem{braak_symm} Braak D. 2019 Symmetries in the quantum Rabi model. \textit{Symmetry} \textbf{11}, 1259.

\bibitem{graham} Graham R, Schlautmann M, and Zoller P. 1992 Observation of Dynamical Localization in Atomic Momentum Transfer: A New Testing Ground for Quantum Chaos. \textit{Phys. Rev. A } \textbf{45}, R19. 

\bibitem{BRODIER200288} Brodier O, Schlagheck P, and Ullmo D. 2002 Resonance-Assisted Tunneling. \textit{Ann. Phys.} \textbf{300}, 88. 

\bibitem{PhysRevLett.115.104101} Gehler Stefan, L\"ock Steffen,  Shinohara S, B\"acker Arnd, Ketzmerick Roland, Kuhl Ulrich, and St\"ockmann Hans-J\"urgen. 2015 Experimental Observation of Resonance-Assisted Tunneling. \textit{Phys. Rev. Lett. } \textbf{115}, 104101. 

\bibitem{stockmann_1999} Stöckmann Hans-Jürgen. 1999  \textit{Quantum Chaos: An Introduction}. Cambridge, UK: Cambridge University Press.

\bibitem{gba} Gaspard P, Alonso D, Burghardt I. 1995 New ways of understanding semiclassical quantization. \textit{Advances in Chemical Physics} 105. 

\bibitem{jain2004} Jain S R. 2004 Semiclassical deuteron. \textit{Journal of Physics G: Nuclear and Particle Physics} \textbf{30}, 157.

\bibitem{mott} Mott, N F, 1961 The theory of impurity conduction, Adv. Phys. textit{10}, 107. 

\bibitem{braak_nl} Ying Z J, Felicetti S, Liu G, Braak D. 2022 Critical quantum metrology in the non-linear quantum Rabi model. \textit{Entropy} \textbf{24}, 1015.
 
\bibitem{krjj} Kumari K, Rajpoot G, Joshi S, and Jain S R. 2023 Qubit control using quantum Zeno effect: action principle approach. \textit{Ann. Phys.} \textbf{450}, 169222. 

\bibitem{ssj} Saini R K, Sehgal R, and Jain S R. 2022 Protection of qubits by nonlinear resonances. \textit{Eur. Phys. J. Plus} \textbf{137}, 356. 


\end{thebibliography}
\end{document}